\documentclass[12pt,a4paper]{article}
\usepackage[
    a4paper,
    top=2.5cm,
    bottom=2.5cm,
    left=2cm,
   right=2cm
]{geometry}
\usepackage{fontspec}
\usepackage{unicode-math}
\usepackage{amsmath}
\usepackage{mathtools}
\usepackage{microtype}
\usepackage{textcomp}
\usepackage{graphicx}
\usepackage{booktabs}
\usepackage{array}
\usepackage{enumitem}
\usepackage{titlesec}
\titleformat{\section}
  {\normalfont\bfseries\large}
  {\thesection.}
  {0.5em}
  {}
\titleformat{\subsection}
  {\normalfont\bfseries\normalsize}
  {\thesubsection.}
  {0.5em}
  {}
\usepackage{setspace}
\usepackage[
    hidelinks
]{hyperref}

\newcommand{\dd}{\mathop{}\!\mathrm{d}}

\begin{document}

\begin{center}

{\Large\bfseries
Nonintegrability theorem on perturbed integrable Hamiltonian\\
dynamical systems with two independent resonances
\par}

\vspace{1em}

Zafeiriadis Nikos$^{1}$ \qquad
Meletlidou Efthymia$^{2}$ \qquad
Efthymiopoulos Christos$^{3}$ \qquad
Volos Christos$^{4}$

\end{center}

\vspace{0.2em}

\begin{flushleft}
\textit{
\small
\noindent
1. Department of Physics, Aristotle University of Thessaloniki 54124, Greece\\
2. Laboratory of Theoretical Mechanics and Astrodynamics, Section of Astrophysics Astronomy and Mechanics, Department of Physics, Aristotle University of Thessaloniki 54124, Greece\\
3. Laboratory of Theoretical Mechanics and Astrodynamics, Section of Astrophysics Astronomy and Mechanics, Department of Physics, Aristotle University of Thessaloniki 54124, Greece\\
4. Laboratory of Nonlinear Systems, Circuits \& Complexity (LaNSCom), Department of Physics, Aristotle University of Thessaloniki, Thessaloniki 54124, Greece
}
\end{flushleft}

\vspace{0.2em}

\begin{abstract}
In this paper we establish a new theorem for nonintegrability of a perturbed,
autonomous, two-degree-of-freedom Hamiltonian dynamical system with two
independent resonances, based on a criterion which utilizes the well-known
theorem of Poincaré and a Lie-series symplectic transformation. We also make
certain assumptions and suggestions as to when the theorem holds and when
it does not.
\end{abstract}
\begin{flushleft}

\section{Introduction}

A very important qualitative feature of Hamiltonian systems is integrability [8].
According to the Liouville--Arnold theorem([1], p.~270--272), if an autonomous Hamiltonian
system with $n$ degrees of freedom possesses $n$ functionally independent
first integrals in involution, then every regular, compact, and connected
common level set of these integrals is diffeomorphic to an invariant
$n$-dimensional torus $\mathbb{T}^n$. Moreover, in a neighbourhood of
such a torus, action--angle coordinates can be introduced.

Let us begin with a perturbed, autonomous, two-degree-of-freedom system given
by a Hamiltonian of the general form. Given that the perturbed Hamiltonian is
an analytic function in terms of the parameter $\varepsilon$, on an open
interval around $\varepsilon=0$, it can consequently be expressed as a power
series in $\varepsilon$ ([2], p.~40),
\end{flushleft}
\begin{equation}
H(J_1,J_2,w_1,w_2)=H_0(J_1,J_2)+\varepsilon H_1(J_1,J_2,w_1,w_2)
\tag{1}
\end{equation}
\begin{flushleft}

The perturbed Hamiltonian is written in curvilinear symplectic coordinates
called action-angle variables in an open region of the phase space:
\end{flushleft}
\[
J_i\in S\subset\mathbb{R}^{n},
\qquad
w_i \bmod 2\pi\in\mathbb{T}^{n}
\]
\begin{flushleft}

More specifically, $H_1(J_1,J_2,w_1,w_2)$ is the perturbation, which is
periodic in $w_1,w_2$ and can be expanded as a double Fourier series with
period $2\pi$. The function $H_0(J_1,J_2)$ is the Hamiltonian of the
unperturbed problem. It is Liouville--Arnold integrable, with finite motions;
that is, in the corresponding regular open region of phase space, the unperturbed
system is locally foliated by invariant $n$-dimensional tori. These tori are invariant with respect to the phase flow of the unperturbed system. Furthermore,
$H_0$ satisfies the nondegeneracy condition ([2], p.~40):
\end{flushleft}
\[
\det
\left(
\frac{\partial^{2}H_0}
{\partial J_i\partial J_j}
\right)
\neq 0,
\qquad
i,j=1,2
\]
\begin{flushleft}
The motion of a phase point is defined by the solution of the system
\end{flushleft}

\begin{equation}
\dot{J}_i=0,
\qquad
\dot{w}_i=\omega_i(J),
\qquad
\omega_i(J)=\frac{\partial H_0}{\partial J_i},
\qquad
i=1,2
\tag{2}
\end{equation}
\begin{flushleft}
That is, 
\end{flushleft}

\begin{equation}
J_i=J_{i0},
\qquad
w_i=\omega_i t+w_{i0},
\qquad
i=1,2
\tag{3}
\end{equation}

\begin{flushleft}
Each torus is determined by the value $J_i=J_{i0}$, while
$w_i\bmod 2\pi$ are the angular coordinates .
\end{flushleft}

\begin{flushleft}
In an abstract $n$-dimensional system, a torus $\mathbb{T}^n$ with frequencies
\end{flushleft}
\[
\mathbf{ω}=(\omega_1,\ldots,\omega_n)
\]
\begin{flushleft}
is called non-resonant if~ $\mathbf{k}\cdot\textbf{ω}\neq0
~~\forall \mathbf{k}\in\mathbb{Z}^{n}\setminus\{\mathbf{0}\},
$
\end{flushleft}
\begin{flushleft}
and resonant if there exists a nonzero integer vector
\end{flushleft}
\[
\mathbf{k}\in\mathbb{Z}^{n}\setminus\{\mathbf{0}\}
\]
\begin{flushleft}
such that    
\end{flushleft}
\[
\mathbf{k}\cdot\boldsymbol{\omega}=0
\]
\begin{flushleft}
Obviously, in a two-dimensional system, a torus $\mathbb{T}^{2}$ is said
to be resonant if
\end{flushleft}
\[
\exists\,
(k_1,k_2)\in\mathbb{Z}^{2}\setminus\{(0,0)\}
:
\quad
k_1\omega_1+k_2\omega_2=0
\]
\begin{flushleft}
For $k_1\neq0$ and $\omega_2\neq0$, this gives    
\end{flushleft}
\[
\frac{\omega_1}{\omega_2}
=
-\frac{k_2}{k_1}
\in\mathbb{Q}
\]
\begin{flushleft}

Generally, the nondegenerate condition corresponds to the case in which the
frequencies are functionally independent. In nondegenerate systems the
non-resonant tori form an everywhere dense set of full measure. The resonant
tori form a set of measure zero, which is also everywhere dense
([3], p.~273).
\end{flushleft}
\begin{flushleft}
A phase trajectory on a non-resonant torus covers the torus densely. One can
simply say that when the frequency ratio
\end{flushleft}
\[
\frac{\omega_1}{\omega_2}\in\mathbb{Q},
\]
\begin{flushleft}
the invariant tori contain periodic orbits that close after a finite time,
whereas when the frequency ratio is irrational,    
\end{flushleft}
\[
\frac{\omega_1}{\omega_2}
\in
\mathbb{R}\setminus\mathbb{Q},
\]
\begin{flushleft}
the invariant tori have quasi-periodic orbits that cover the tori densely.
\end{flushleft}

\section{Poincaré's theorem on nonintegrability}

Obviously,
\[
H(J_1,J_2,w_1,w_2)
\]
\begin{flushleft}
is an integral of motion. Let us suppose that $H$ has a second integral of
motion    
\end{flushleft}
\[
\Phi(J_1,J_2,w_1,w_2,\varepsilon)
\]
\begin{flushleft}
in an open neighborhood of $J_1,J_2$, for $w_1,w_2$ from $0$ to $2\pi$,
and for $\varepsilon$ in a neighborhood of $\varepsilon=0$. Then $\Phi$
can be expanded in a convergent series around $\varepsilon=0$:    
\end{flushleft}
\begin{equation}
\Phi
=
\Phi_0(J_1,J_2)
+
\varepsilon\Phi_1(J_1,J_2,w_1,w_2)
+
\varepsilon^2\Phi_2
+
\mathcal{O}(\varepsilon^3)
\tag{4}
\end{equation}
\begin{flushleft}
If we take a non-resonant torus, note that $\Phi_0$ is independent of the
angles. A detailed explanation can be found in ([3] p.~353).

Poincaré proves that $\Phi_0$ is an integral of $H_0$, independent of it.
Furthermore, up to order $\varepsilon$, since $\Phi$ is an integral of $H$,
it holds that
\end{flushleft}
\[
[\Phi,H]=0
\]
\begin{flushleft}
That leads to    
\end{flushleft}
\begin{equation}
[\Phi_0,H_0]=0,
\qquad
[\Phi_0,H_1]+[\Phi_1,H_0]=0
\tag{5}
\end{equation}
\begin{flushleft}
The first relation in (5) states that $\Phi_0$ is an integral of $H_0$.
We shall examine whether $\Phi_0$ is independent of $H_0$. The second
relation in (5) yields the relation between the Fourier coefficients
$B_{k_1,k_2}$ and $A_{k_1,k_2}$:    
\end{flushleft}

\begin{equation}
B_{k_1,k_2}
\left(
k_1\frac{\partial H_0}{\partial J_1}
+
k_2\frac{\partial H_0}{\partial J_2}
\right)
=
A_{k_1,k_2}
\left(
k_1\frac{\partial\Phi_0}{\partial J_1}
+
k_2\frac{\partial\Phi_0}{\partial J_2}
\right)
\tag{6}
\end{equation}
\begin{flushleft}
Here,    
\end{flushleft}
\[
H_1
=
\sum_{k_1,k_2}
A_{k_1,k_2}(J_1,J_2)
\exp\left(
i(k_1w_1+k_2w_2)
\right)
\]
\begin{flushleft}
and
\end{flushleft}
\[
\Phi_1
=
\sum_{k_1,k_2}
B_{k_1,k_2}(J_1,J_2)
\exp\left(
i(k_1w_1+k_2w_2)
\right)
\]
\begin{flushleft}
are the Fourier series of $H_1$ and $\Phi_1$ in $w_1$ and $w_2$.

Relation (6) is an identity on the phase space; that is, it holds for every
$J_i,w_i$ in an open region $d\times\mathbb{T}^{n}$, at least
([2] p.~47).

According to equation (6), its left-hand side becomes zero on the dense set
of resonant tori of the unperturbed system $H_0$, since it is nondegenerate:
    
\end{flushleft}

\begin{equation}
\frac{
\dfrac{\partial H_0}{\partial J_1}
}{
\dfrac{\partial H_0}{\partial J_2}
}
=
\frac{\omega_1}{\omega_2}
=
-\frac{k_2}{k_1}
\tag{7}
\end{equation}
\newpage
\begin{flushleft}
If an infinite set of Fourier coefficients
$A_{k_1,k_2}\neq0$ of $H_1$ are different from zero on these dense resonant
tori, then according to equation (6),    
\end{flushleft}

\begin{equation}
\frac{
\dfrac{\partial\Phi_0}{\partial J_1}
}{
\dfrac{\partial\Phi_0}{\partial J_2}
}
=
-\frac{k_2}{k_1}
\tag{8}
\end{equation}
\begin{flushleft}
If we combine (7) and (8), we get
\end{flushleft}

\begin{equation}
D(J)
=
\frac{\partial(\Phi_0,H_0)}
{\partial(J_1,J_2)}
=
0
\tag{9}
\end{equation}
\begin{flushleft}
Therefore, $\Phi_0$ is functionally dependent on $H_0$, which is a
contradiction. We point out that (9) holds for the entire domain of the
phase space. Since $D$ is continuous, if it is zero on a dense set, then
it is zero everywhere.

Indeed, take any $J^{*}$ in the domain. Since the resonant set of tori is
dense, there exists a sequence
    
\end{flushleft}
\[
J^{(n)}\to J^{*}
\]
\begin{flushleft}
such that
\end{flushleft}
\[
D\left(J^{(n)}\right)=0.
\]
\begin{flushleft}
Therefore,
\end{flushleft}
\[
D(J^{*})
=
\lim_{n\to\infty}
D\left(J^{(n)}\right)
=
0,
\]
\begin{flushleft}
so (9) holds throughout the entire domain.
Thus, Poincaré's theorem [7] states:
\end{flushleft}
\medskip
\noindent
\textit{If an infinite set of Fourier coefficients
$A_{k_1,k_2}\neq0$ of $H_1$ are nonzero on a dense set of resonant tori of
the unperturbed, integrable, nondegenerate system, then $H$ does not possess
a second integral of motion analytic in the real variables in an open
neighborhood of $J_1,J_2$, for $w_1,w_2$ from $0$ to $2\pi$, and for
$\varepsilon$ in a neighborhood of $\varepsilon=0$.}


\section{Meletlidou--Ichtiaroglou nonintegrability criterion and equivalence to Poincaré's theorem on nonintegrability}
\begin{flushleft}
A more detailed presentation is given in [4].
From the second relation in (5), we obtain    
\end{flushleft}
\[
[\Phi_0,H_1]
=
[H_0,\Phi_1],
\]
\begin{flushleft}
and therefore
\end{flushleft}
\begin{equation}
\frac{\dd\Phi_1}{\dd t}
=
[\Phi_0,H_1]
\tag{10}
\end{equation}
\begin{flushleft}
Consider a member of the family of resonant tori for which
\end{flushleft}

\[
\frac{\omega_1}{\omega_2}
=
\frac{n}{m},
\qquad
n,m\in\mathbb{Z}\setminus\{0\}.
\]
\begin{flushleft}
By integrating (10) over one period, we obtain the following.
\end{flushleft}
\newpage
\begin{flushleft}
The right-hand side of this equation gives zero upon integration, since the
corresponding term is the time derivative of $\Phi_1$ integrated over one
period $T$ on the resonant torus.
\end{flushleft}
\begin{flushleft}
The period $T$ is determined by
\end{flushleft}
\[
T
=
\frac{2\pi}{\Omega},
\qquad
\Omega
=
\frac{\omega_1}{n}
=
\frac{\omega_2}{m}
\]
\begin{flushleft}
The left-hand side results in
\end{flushleft}
\begin{equation}
\left(
k_2\frac{\partial\Phi_0}{\partial J_1}
+
k_1\frac{\partial\Phi_0}{\partial J_2}
\right)
\frac{
\partial\displaystyle\int_0^T H_1\,\dd t
}{
\partial\varphi
}
=
0
\tag{11}
\end{equation}
\begin{flushleft}
where
\end{flushleft}
\[
\varphi
=
k_2w_1+k_1w_2
\]
\begin{flushleft}
We note that every orbit on a resonant torus is periodic and differs from
the others only in terms of $w_{10},w_{20}$. Thus, we obtain an entire
family of periodic orbits on a specific resonant torus.

Along the unperturbed motion,

\end{flushleft}

\[
\frac{\dd\varphi}{\dd t}
=
k_2\dot{w}_1+k_1\dot{w}_2
=
0.
\]
\begin{flushleft}
Therefore, $\varphi$ is constant on every periodic orbit. On a resonant
torus of $H_0$ there are infinitely many periodic orbits parametrized by
$\varphi$.

The mean value of $H_1$ is
\end{flushleft}
\[
\left\langle H_1\right\rangle(\varphi)
=
\frac{1}{T}
\int_0^T
H_1(t;\varphi)\,\dd t
\]
\begin{flushleft}
It can either be the same for all periodic orbits of the torus, or it can
vary as $\varphi$ changes.

In the second case, there exists a trajectory for which

\end{flushleft}
\[
\frac{
\partial\displaystyle\int_0^T H_1\,\dd t
}{
\partial\varphi
}
\neq0
\]
\begin{flushleft}
On a resonant torus, this leads to
\end{flushleft}
\[
\frac{
\dfrac{\partial\Phi_0}{\partial J_1}
}{
\dfrac{\partial\Phi_0}{\partial J_2}
}
=
-\frac{k_1}{k_2}
\]
\begin{flushleft}
This relation concerns the entire resonant torus.

It means that $\Phi_0$ depends on $H_0$ on this torus. This occurs for a
dense set of resonant tori if

\end{flushleft}
\[
\frac{
\partial\displaystyle\int_0^T H_1\,\dd t
}{
\partial\varphi
}
\neq0
\]
\begin{flushleft}
on this dense set. But this is not possible according to Poincaré's theorem,
since $\Phi_0$ is independent of $H_0$. Therefore, $H$ is nonintegrable.

More specifically, the exact proposition of [4] is that if there exists a
dense set of resonant tori of $H_0$ such that, on each of them, the average
value of $H_1$ depends on the particular periodic orbit, then the perturbed
Hamiltonian is nonintegrable for $\varepsilon
\in
(-\varepsilon_0,\varepsilon_0)\setminus\{0\}$ in an open neighborhood of $\varepsilon=0$.
\end{flushleft}
\begin{flushleft}
It is proven that the Meletlidou--Ichtiaroglou criterion and Poincaré's
theorem are equivalent.
\end{flushleft}


\section{A new theorem on nonintegrability in the case of two independent resonances}
\begin{flushleft}
We are going to establish a new theorem on the nonintegrability of a
perturbed Hamiltonian system with two resonances.

We begin with the Hamiltonian

\end{flushleft}
\begin{equation}
H
=
H_0(J_1,J_2)
+
\varepsilon
\left[
A_1(J_1,J_2)
\cos(k_1w_1-k_2w_2)
+
A_2(J_1,J_2)
\cos(l_1w_1-l_2w_2)
\right]
\tag{12}
\end{equation}
\begin{flushleft}
Here $J_1,J_2,w_1,w_2$ are the action-angle variables, $H_0$ is the
integrable, nondegenerate, unperturbed part in action-angle variables, and
$H_1$ is the perturbation with two resonances, $k_1,k_2,l_1,l_2\in\mathbb{Z}.$ and arbitrary.

The vectors
\end{flushleft}
\[
\mathbf{k}=(k_1,-k_2),
\qquad
\mathbf{l}=(l_1,-l_2)
\]
\begin{flushleft}
are independent.

We suppose, as in Poincaré's theorem of nonintegrability, that there exists
a second integral of motion independent of $H$, and we are going to use
Poincaré's lemma that $\Phi_0$ is independent of $H_0$.

To study the nonintegrability generated by the interaction of these two
resonances, we apply a sequence of Lie-series symplectic transformations ([6]).
\end{flushleft}
\begin{flushleft}
The Lie operator is defined by the Poisson bracket:

\end{flushleft}
\[
L_WF
=
[H,F].
\]
\begin{flushleft}
Our Lie generators $W_{m_k,n_k}$ is a sequence of function in angle variables
\end{flushleft}
\begin{equation}
W_{m_k,n_k}
=
\cos(m_kw_1-n_kw_2),
\tag{14}
\end{equation}
\begin{flushleft}
where    
\end{flushleft}
\[
m_k=p_kk_1+q_kl_1,
\qquad
n_k=j_kk_2+s_kl_2,
\]
\begin{flushleft}
with 
\end{flushleft}
\[
p_k,q_k,j_k,s_k,k_1,k_2,l_1,l_2,\in\mathbb{Z}-\{0\},~~k\in\mathbb{N}-\{0\}
\qquad
i=1,2,
\]
\begin{flushleft}
Then the transformed Hamiltonian is
\end{flushleft}
\begin{equation}
\begin{aligned}
H'
&= \exp\!\left(\varepsilon L_{W_{m_k,n_k}}\right)H
 = \sum_{n=0}^{\infty}
   \frac{\varepsilon^n}{n!}L_{W_{m_k,n_k}}^n H=
\\[4pt]
&= H_0 + \varepsilon H_1
 + \varepsilon [W_{m_k,n_k},H_0]
 + \varepsilon^2 [W_{m_k,n_k},H_1]
 + \frac{\varepsilon^2}{2}[W_{m_k,n_k},[W_{m_k,n_k},H_0]]
 + \mathcal{O}(\varepsilon^3).
\end{aligned}
\tag{15}
\end{equation}
\begin{flushleft}
According to Cauchy's theorem for Lie series, the series for
$H'$ converges, since it does not have any small divisors[5].

At each step we apply $W_{m_k,n_k}$ to the initial Hamiltonian $H$. At each step we shall prove that eq.(17) is non-zero at the resonant torus with $\frac{ω_1}{ω_2}=\frac{k_2+n_k}{k_1+m_k}$.
After applying this infinite sequence of Lie generators, we shall prove that (17)  is non-zero at the resonant tori $\frac{ω_1}{ω_2}=\frac{k_2+n_k}{k_1+m_k}$ where $p_k,q_k,j_k,s_k$ take infinite values and $\frac{\omega_1}{\omega_2} $ at these resonant tori form a dense set in action space.
\newpage
We also apply the Poisson bracket with $W_{m_k,n_k}$ to equation (5):

\end{flushleft}
\[
[W_{m_k,n_k},[\Phi_0,H_1]]
+
[W_{m_k,n_k},[\Phi_1,H_0]]
=
0
\]
\begin{flushleft}
Using again Jacobi's identity for
\end{flushleft}
\[
[W_{m_k,n_k},[\Phi_0,H_1]]
=
[\Phi_0,[H_1,W_{m_k,n_k}]]
+
[[W_{m_k,n_k},\Phi_0],H_1]
\]
\begin{flushleft}
and
\end{flushleft}
\[
[W_{m_k,n_k},[\Phi_1,H_0]]
=
[\Phi_1,[W_{m_k,n_k},H_0]]
+
[[W_{m_k,n_k},\Phi_1],H_0],
\]
\begin{flushleft}
we obtain
\end{flushleft}
\begin{equation}
[\Phi_0,[W_{m_k,n_k},H_1]]+[[W_{m_k,n_k},\Phi_0],H_1]+[\Phi_1,[W_{m_k,n_k},H_0]]+[[W_{m_k,n_k},\Phi_1],H_0]=0
\tag{16}
\end{equation}
\begin{flushleft}
According to the criterion by E.~Meletlidou and S.~Ichtiaroglou, we
integrate (16) on the periodic orbits of the unperturbed invariant resonant
tori of $H_0$, where the resonance condition holds:

\end{flushleft}
\[
\frac{\omega_1}{k_2+n_k}
=
\frac{\omega_2}{k_1+m_k}
\tag{$\alpha$}
\]
\begin{flushleft}
And also
\end{flushleft}
\[
\frac{\omega_1}{l_2+n_k}
=
\frac{\omega_2}{l_1+m_k}
\tag{$\beta$}
\]
\begin{flushleft}
The results of the analysis using $(\alpha)$ also hold for $(\beta)$.
Let us consider the resonance condition $(\alpha)$.

The period $T$ is

\end{flushleft}
\[
T
=
\frac{2\pi}{\Omega},
\qquad
\Omega
=
\frac{\omega_1}
{k_2(1+j_k)+s_kl_2}
=
\frac{\omega_2}
{k_1(1+p_k)+q_kl_1}
\]
\begin{flushleft}
We integrate over one period:   
\end{flushleft}
\[
\int_0^T
\left(
[W_{m_k,n_k},[\Phi_0,H_1]]
+
[W_{m_k,n_k},[\Phi_1,H_0]]
\right)\dd t
\]
\begin{flushleft}
Thus,  
\end{flushleft}
\begin{equation}
\int_0^T
\Big(
[\Phi_0,[W_{m_k,n_k},H_1]]
+
[[W_{m_k,n_k},\Phi_0],H_1]
+
[\Phi_1,[W_{m_k,n_k},H_0]]
+
[[W_{m_k,n_k},\Phi_1],H_0]
\Big)
\dd t=0
\tag{17}
\end{equation}
\begin{flushleft}
Consider  
\end{flushleft}
\[
w_1
=
\omega_1t+w_{10},
\qquad
w_2
=
\omega_2t+w_{20},
\]
\begin{flushleft}
with frequencies of the form
\end{flushleft}
\[
\omega_1
=
\left[
k_2(1+j_k)+s_kl_2
\right]\Omega,
\]

\[
\omega_2
=
\left[
k_1(1+p_k)+q_kl_1
\right]\Omega
\]
\newpage
\begin{flushleft}
For the first term of eq.(17) which is 
\[
\int_0^T[\Phi_0,[W,H_1]]dt  
\]
and selecting the term $\cos((k_1+m_1)w_1-(k_2+m_2)w_2)$ as a general term, we obtain
\end{flushleft}
\[
\begin{aligned}
&\frac{1}{2}
\left(
\frac{\partial A_1}{\partial J_2}\,n_k
-
\frac{\partial A_1}{\partial J_1}\,m_k
\right)
\left(
\frac{\partial \Phi_0}{\partial J_2}(k_2+n_k)
-
\frac{\partial \Phi_0}{\partial J_1}(k_1+m_k)
\right)\cdot
\\[6pt]
&\qquad\cdot
\int_0^T
\cos\!\left(
(k_1+m_k)w_1-(k_2+n_k)w_2
\right)\,dt .
\end{aligned}
\]
 
\begin{flushleft}
Provided that $\Phi_0$ is independent of $H_0$, the term   
\end{flushleft}
\[
(k_2+n_k)
\frac{\partial\Phi_0}{\partial J_2}
-
(k_1+m_k)
\frac{\partial\Phi_0}{\partial J_1}
\] is different from zero
\begin{flushleft}
Also the other coefficient is nonzero if
\end{flushleft}
\[
\frac{n_k}{m_K}
\neq
\frac{
\dfrac{\partial A_1}{\partial J_1}
}{
\dfrac{\partial A_1}{\partial J_2}
}
\]
\begin{flushleft}
as well as    
\end{flushleft}

\begin{flushleft}
The integral is nonzero since,
on the above resonance $(\alpha)$,
\end{flushleft}
\begin{equation}
\begin{aligned}
\int_0^T
&\cos\!\left[(k_1+m_k)w_1-(k_2+n_k)w_2\right]\,dt
=
\int_0^T \cos\phi_k\,dt
=
T\cos\phi_k .
\end{aligned}
\tag{18}
\end{equation}
where $\varphi_k=(k_1+m_k)w_{10}-(k_2+n_k)w_{20}$ since the resonance does not vanish the initial conditions $w_{i0}$.
\begin{flushleft}
Looking at the last term of (17), it can be proven that it is zero, since
it is the time derivative of $[W_{m_k,n_k},\Phi_1]$:
\end{flushleft}
\[
\begin{aligned}
\int_0^T [[W_{m_k,n_k},\Phi_1],H_0]\,dt
&=
\int_0^T
\left(
\frac{\partial [W_{m_k,n_k},\Phi_1]}{\partial w_1}\frac{\partial H_0}{\partial J_1}
+
\frac{\partial [W_{m_k,n_k},\Phi_1]}{\partial w_2}\frac{\partial H_0}{\partial J_2}
\right)dt=
\\[6pt]
&=
\int_0^T
\left[
\frac{\partial [W_{m_k,n_k},\Phi_1]}{\partial w_1}\dot{w}_1
+
\frac{\partial [W_{m_k,n_k},\Phi_1]}{\partial w_2}\dot{w}_2
\right]dt .
\end{aligned}
\]
\begin{flushleft}
That is,    
\end{flushleft}
\[
\int_0^T
\frac{\dd[W_{m_k,n_k},\Phi_1]}{\dd t}
\,\dd t
=
0
\]
\begin{flushleft}
The other terms are   
\end{flushleft}
\[
\int_0^T
\left(
[[W_{m_k,n_k},\Phi_0],H_1]
+
[\Phi_1,[W_{m_k,n_k},H_0]]
\right)
\dd t
\tag{19}
\]
\begin{flushleft}
The second term is 
\end{flushleft}
\[
\int_0^T
[[W_{m_k,n_k},\Phi_0],H_1]
\,\dd t
\]
\begin{flushleft}
and has the coefficient
\end{flushleft}
\begin{equation*}
\begin{aligned}
\frac{1}{2}\Bigg[
&\frac{\partial A_1}{\partial J_1}\,m_k
\left(
\frac{\partial \Phi_0}{\partial J_2}\,n_k
-
\frac{\partial \Phi_0}{\partial J_1}\,m_k
\right)
+
\frac{\partial A_1}{\partial J_2}\,n_k
\left(
\frac{\partial \Phi_0}{\partial J_1}\,m_k
-
\frac{\partial \Phi_0}{\partial J_2}\,n_k
\right)
\\[6pt]
&+
A_1
\left(
-\frac{\partial^2\Phi_0}{\partial J_1\partial J_2}
\,k_1k_2(p_k+j_k)
+k_1l_2s_k+k_2l_1q_k
+k_1\frac{\partial^2\Phi_0}{\partial J_1^2}\,m_k
+k_2\frac{\partial^2\Phi_0}{\partial J_2^2}\,n_k
\right)
\Bigg]
\end{aligned}
\end{equation*}
\begin{flushleft}
It can be proven that this it is nonzero when $H_0$ does not have a linear
integral in $J_1,J_2$, namely $\Phi_0$, that is continued in the perturbed
system. Also, the integral (18) and the last term in (19) at the resonance is nonzero.

Also, the terms do not balance.

We shall prove in Appendix A. that the ratios

\end{flushleft}

\[
\frac{\omega_1}{\omega_2}
=
\frac{k_2+n_k}{k_1+m_k}
\]
\begin{flushleft}
and
\end{flushleft}
\[
\frac{\omega_1}{\omega_2}
=
\frac{l_2+n_k}{l_1+m_k}
\]
\begin{flushleft}
form a dense set in the relevant interval. Therefore, the resonant tori form a
dense set in the phase space. After proving that the resonant tori are dense in the action space, the Hamiltonian given by (12) is nonintegrable

For \(H\) to possess a second independent first integral analytic in \(\varepsilon\), equation (17) must vanish. Hence, the vanishing of (17) is a necessary condition for the existence of such an integral. Under the assumptions stated below, equation (17) is non-zero on a dense set of resonant tori, which yields a contradiction.

Finally,according to the criterion, the Hamiltonian system given by (12) is
nonintegrable, except in the cases described above.

We formulate the following theorem:
\end{flushleft}
\begin{flushleft}
\medskip
\noindent
\textbf{Theorem:}
Consider the analytic autonomous Hamiltonian system
\[
H(J,w;\varepsilon)
=
H_0(J_1,J_2)
+
\varepsilon
\left[
A_1(J_1,J_2)\cos(k_1w_1-k_2w_2)
+
A_2(J_1,J_2)\cos(l_1w_1-l_2w_2)
\right],
\]
where $H_0$ is Liouville-integrable and non-degenerate, and
\[
\mathbf{k}=(k_1,-k_2),
\qquad
\mathbf{l}=(l_1,-l_2)
\]
are linearly independent integer vectors.

Let
\[
W_{m_k,n_k}
=
\cos(m_kw_1-n_kw_2),
\]
with
\[
m_k=p_kk_1+q_kl_1,
\qquad
n_k=j_kk_2+s_kl_2,
\]
where
\[
p_k,q_k,j_k,s_k\in\mathbb{Z}-\{0\},
\qquad
k\in\mathbb{N}-\{0\}.
\]

Assume that the corresponding resonance ratios
\[
\frac{\omega_1}{\omega_2}
=
\frac{k_2+n_k}{k_1+m_k}
\]
or, analogously,
\[
\frac{\omega_1}{\omega_2}
=
\frac{l_2+n_k}{l_1+m_k},
\]
generate a dense set of resonant tori in the considered action domain.

Assume further that, on this dense resonant set, the second-order equation (17) does not vanish.
In particular, either for the relevant resonant contribution,
\[
n_k\frac{\partial A_1}{\partial J_2}
-
m_k\frac{\partial A_1}{\partial J_1}
\neq 0,
\]
with analogous non-vanishing condition for $A_2$, or $H_0$ does not possess a linear integral with respect to $J_1,J_2$ or that the
remaining averaged Poisson-bracket contributions do not cancel each other we proved that for sufficiently small
\[
\varepsilon\neq 0,
\]
the perturbed Hamiltonian $H$ admits no second first integral which is
functionally independent of $H$ and analytic in $\varepsilon$ in the
considered open domain that is continued in the perturbed system. Consequently $H$ is nonintegrable.

\textbf{Remark:}

The theorem is inconclusive in cases:

1)Where all second-order resonant coefficient vanishes,as well as the coefficients,
    \[
    n_k\frac{\partial A_i}{\partial J_2}
    -
    m_k\frac{\partial A_i}{\partial J_1}
    =0,~i=1,2,k\in\mathbb{N}-\{0\}
    \]

2)The two Fourier vectors are not independent, i.e.
    \[
    \mathbf{k}\parallel\mathbf{l};
    \]
then it is known that the system is integrable.

\end{flushleft}
Failure of any of these hypotheses does not imply integrability;it only means that the present criterion is inconclusive.
\begin{flushleft}
\section{Conclusions}

In this work we have investigated a two-degree-of-freedom nearly
integrable Hamiltonian system whose perturbation contains two linearly
independent Fourier modes. By combining a Lie-series symplectic
transformation with the non-integrability approach of Poincaré and the
criterion of Meletlidou--Ichtiaroglou, we obtain a second-order resonant
obstruction to the existence of an additional analytic first integral.

The Lie transformation generates combination harmonics at order
$\mathcal{O}(\varepsilon^2)$. When one of these harmonics becomes resonant
with an unperturbed invariant torus, its average over the corresponding
periodic orbit does not, in general, vanish and depends on the phase
labelling the periodic orbit on that torus.

If this non-vanishing, orbit-dependent averaged contribution occurs on a
dense set of resonant tori, and if the remaining second-order averaged
terms do not cancel it, the necessary condition for the existence of a
second first integral analytic in $\varepsilon$ is violated. Consequently,
for sufficiently small $\varepsilon\neq0$, no second first integral
functionally independent of $H$ and analytic in the stated class can
exist in the considered domain.

Thus, the obstruction is not produced merely by the presence of two
independent Fourier modes, but by their second-order interaction together
with the density and non-vanishing conditions stated in the theorem.

The Lie-series formulation is particularly convenient for this analysis
because it preserves the canonical structure and provides an explicit
order-by-order description of the harmonics generated by the interaction
of the perturbing terms.
\end{flushleft}

\appendix
\section{Dense set of Resonant Tori}
\appendix
\label{app:density}
We prove that the resonance ratios
\[
\frac{k_2+n_k}{k_1+m_k}
\]
form a dense set in the relevant positive frequency-ratio interval.

Recall that
\[
m_k=p_k k_1+q_k l_1,
\qquad
n_k=j_k k_2+s_k l_2,
\]
where
\[
p_k,q_k,j_k,s_k\in\mathbb Z\setminus\{0\}.
\]

Let
\[
d_1=\gcd(k_1,l_1),
\qquad
d_2=\gcd(k_2,l_2).
\]
By B\'ezout's identity,
\[
\{~p k_1+q l_1:\;p,q\in\mathbb Z~\}=d_1\mathbb Z,
\]
and similarly,
\[
\{~jk_2+s l_2:\;j,s\in\mathbb Z~\}=d_2\mathbb Z.
\]
Therefore $m_k$ and $n_k$ may attain arbitrary integer
multiples of $d_1$ and $d_2$, respectively.

Consider an arbitrary open interval
\[
(\alpha,\beta)\subset\mathbb R^+,
\qquad
0<\alpha<\beta.
\]

Since $m_k$ may take arbitrarily large values, we can choose
$m_k$ such that
\[
D_k:=k_1+m_k>0
\]
and
\[
\frac{d_2}{D_k}<\beta-\alpha.
\tag{A.1}
\]

Now choose $n_k$ such that
\[
N_k:=k_2+n_k
\]
which is a large enough multiple of $d_2$ satisfying 
\[
N_k>\alpha D_k \Rightarrow \frac{N_k}{D_k}>α
\]
Therefore the set of integers 
\[
Β=\{n_k:k \in \mathbb{N},~~ \frac{N_k}{D_k}>a\}
\]
is nonempty.
In the above set there is a minimum integer $n_k$ such that $S_k=min ~n_k,~~n_k \in B$ for which it holds that $S_k\in B$

We will show that for $S_k$ it holds that
\[
\frac{S_k}{D_k}<β
\]
Let us suppose the opposite
\[
\frac{S_k}{D_k}\geβ
\]
then,
\[
\frac{S_k}{D_k}-\frac{d_2}{D_k}\geβ-\frac{d_2}{D_k}>β-(β-α)=α
\]
which is a contradiction, so 
\[
\frac{S_k}{D_k}<β
\]
since $S_k$ is the minimum of the set $B$, and $D_k$ may attain arbitrary large values,
\[
\boxed{α<\frac{S_k}{D_k}<β}
\]
Τhe interval $(\alpha,\beta)$ is arbitrary, the set
\[
\left\{
\frac{k_2+n_k}{k_1+m_k}
\right\}
\]
is dense in $\mathbb R^+$.

The same argument applies to the second family
\[
\left\{
\frac{l_2+n_k}{l_1+m_k}
\right\},
\]
and therefore it is also dense in $\mathbb R^+$.

Finally, since the unperturbed Hamiltonian $H_0$ is non-degenerate,
the map
\[
J\mapsto
\omega(J)=\nabla H_0(J)
\]
is locally invertible. Hence the the above rational
frequency ratios imply that the corresponding resonant invariant
tori form a dense set in the considered regular action domain.
\label{app:density}
\newpage
\section*{References}

\begin{enumerate}[label={[\arabic*]}]

\item
Arnold, V.~I. (1989).
\textit{Mathematical Methods of Classical Mechanics}.
2nd ed.
New York: Springer.

\item
Meletlidou, E. (1995).
\textit{Non-Integrability Criteria in Hamiltonian Dynamical Systems}.
PhD dissertation, Aristotle University of Thessaloniki.

\item
Arnold, V.~I., Kozlov, V.~V., and Neishtadt, A.~I. (2006).
\textit{Mathematical Aspects of Classical and Celestial Mechanics}.
3rd ed.
Berlin: Springer.

\item
Meletlidou, E. and Ichtiaroglou, S. (1994).
``A criterion for non-integrability based on Poincaré's theorem.''
\textit{Physica D: Nonlinear Phenomena},
71(3), 261--268.
ISSN 0167-2789.

\url{https://doi.org/10.1016/0167-2789(94)90148-1}

\item
Giorgilli, A. (2002).
``Notes on Exponential Stability of Hamiltonian Systems.''
Centro di Ricerca Matematica Ennio De Giorgi.

\item
Hori, G.-I. (1966).
``Theory of General Perturbations with Unspecified Canonical Variables.''
\textit{Publications of the Astronomical Society of Japan},
18(4), 287--296.
\item
H.Poincaré, 1899, “Les Méthodes Nouvelles de la Méchanique Céleste", Vol III Gauthier-
Villars, Paris, (“New Methods of Celestial Mechanics”, Vol.III, Ed. D.Goroff).

\item
Gaetano Vilasi, 2001, “Hamiltonian Dynamics”, World Scientific.
\end{enumerate}

\end{document}